\documentclass{article}

\usepackage{amssymb}
\usepackage{amsmath}
\usepackage{epsfig}

\begin{document}

\title{Cosmologies with Energy Exchange}
\author{{John D. Barrow\thanks{
e-mail: J.D.Barrow@damtp.cam.ac.uk} and T. Clifton\thanks{
e-mail: T.Clifton@damtp.cam.ac.uk}} \\
{\small {\textit{Department of Applied Mathematics and Theoretical Physics,}}}\\
{\small {\textit{University of Cambridge, Cambridge CB3 0WA, UK}}}}
\date{{\normalsize {\today}}}
\maketitle

\begin{abstract}
We provide a simple mathematical description of the exchange of energy
between two fluids in an expanding Friedmann universe with zero spatial
curvature. The evolution can be reduced to a single non-linear differential
equation which we solve in physically relevant cases and provide an analysis
of all the possible evolutions. Particular power-law solutions exist for the
expansion scale factor and are attractors at late times under particular
conditions. We show how a number of problems studied in the literature, such
as cosmological vacuum energy decay, particle annihilation, and the
evolution of a population of evaporating black holes, correspond to simple
particular cases of our model. In all cases we can determine the effects of
the energy transfer on the expansion scale factor. We also consider the
situation in the presence of `anti-decaying' fluids and so called `phantom'
fluids which violate the dominant energy conditions.
\end{abstract}

\section{Introduction}

There are many cosmological situations where the transfer of energy between
two fluids is important. The interaction between matter and radiation \cite%
{tol}, the decay of massive particles into radiation \cite{mass}, matter
creation \cite{create}, the formation and evaporation of primordial black
holes \cite{pbh}, the annihilations of particle-antiparticle pairs \cite{ann}%
, particle or string production \cite{tur, jb}, inflaton decay \cite{warm}
and the decay of some scalar field \cite{scal} or vacuum energy \cite{free},
are all particular examples which have been studied in general-relativistic
cosmology. The situation in Brans-Dicke cosmology has also been investigated 
\cite{bd}, as have the cases of two arbitrary interacting fluids \cite{arb}
and more than two interacting fluids \cite{three}. In some cases, as in the
example of accreting and evaporating black holes, there will be a two-way
transfer of energy occurring as, say, a spectrum of radiation
inhomogeneities collapse under their self-gravity in the early universe to
form a population of primordial black holes but the products of the Hawking
evaporation of the black holes adds to the cosmological population of
interacting relativistic particles \cite{pbh}. The different studies of
these particular situations have often identified the existence of special
power-law scaling solutions. In this paper we consider a general problem of
this sort, describe its general behaviour, relate it to the existence of
special power-law solutions, and describe its general solution succinctly in
terms of the parameters defining the energy exchanges. The examples in the
literature can then be shown to be particular examples of these solutions
and the conditions for their stability are made clear.

We will consider the mutual exchange of energy between two fluids at rates
that are proportional to a linear combination of their individual densities
and the expansion rate of the universe. In the absence of any interaction
the fluids reduce to two separate perfect fluids.

\section{Decaying Fluids}

\label{decaysec}

Consider a flat Friedmann Robertson Walker (FRW) universe with expansion scale factor $a(t)$ containing
two fluids with equations of state 
\begin{eqnarray*}
p &=&(\gamma -1)\rho , \\
p_{1} &=&(\Gamma -1)\rho _{1},
\end{eqnarray*}%
where the $\gamma $ and $\Gamma $ are constants, and the evolution of the
Hubble parameter $H=\dot{a}/a$ is governed by the Friedmann equation 
\begin{equation}
3H^{2}=\rho +\rho _{1},  \label{1}
\end{equation}%
where $8\pi G\equiv 1$. Assume that the two fluids exchange energy but the
total energy is conserved so that 
\begin{eqnarray}
\dot{\rho}_{1}+3H\Gamma \rho _{1} &=&-\beta H\rho _{1}+\alpha \rho H,
\label{2} \\
\dot{\rho}+3H\gamma \rho &=&\ \beta H\rho _{1}-\alpha \rho H,  \label{3}
\end{eqnarray}%
where $\alpha $ and $\beta $ are constants parametrising the energy
exchanges between the two fluids. Generalisations of this simple cosmology
to spatially curved or anisotropic universes can be made in an obvious way
if required \cite{jb, jb1}. In an expanding universe ($H>0$) this scenario
corresponds to $\rho $ and $\rho _{1}$ `decaying' into each other in
proportion to their energy densities if $\alpha $ and $\beta $ are positive.
The degenerate case $\gamma =\Gamma $ can be seen to be trivially equivalent
to the standard scenario without energy exchange, by considering the fluid $%
\rho _{2}=\rho +\rho _{1}$.

Using the last three equations we can eliminate the densities to obtain a
single master equation for the Hubble expansion, $H(t):$ 
\begin{equation}
\ddot{H}+H\dot{H}(\alpha +\beta +3\gamma +3\Gamma )+\frac{3}{2}H^{3}(\alpha
\Gamma +\beta \gamma +3\Gamma \gamma )=0.  \label{H}
\end{equation}%
Let us rewrite eq. (\ref{H}) as 
\begin{equation}
\ddot{H}+AH\dot{H}+BH^{3}=0  \label{master}
\end{equation}%
with 
\begin{eqnarray*}
A &\equiv &\alpha +\beta +3\gamma +3\Gamma , \\
B &\equiv &\frac{3}{2}(\alpha \Gamma +\beta \gamma +3\Gamma \gamma ).
\end{eqnarray*}%
This equation is a special case of the more general differential
equation\footnote{$\ddot{y}+\alpha f\dot{y}+\beta \int fdy+\gamma f=0$, where $%
y=y(x),$ $f=f(y)$ and overdots denote differentiation with respect to $x$. $%
\alpha ,\beta $ and $\gamma $ are constants.} considered by Chimento \cite%
{Chim}. In \cite{Chim} Chimento investigates the mathematical structure of
this equation by showing that it has a form invariance, which is
subsequently used to find solutions.

Simple self-similar solutions to eq. (\ref{master}) exist with 
\begin{equation}
H=\frac{h}{t}  \label{h}
\end{equation}%
where, for $h\neq 0$, 
\begin{equation*}
2-Ah+Bh^{2}=0,
\end{equation*}%
and there are two non-trivial solutions, $H_{+}(t)$ and $H_{-}(t)$, with $h$
values 
\begin{equation}
h_{\pm }=\frac{A\pm \sqrt{A^{2}-8B}}{2B}.  \label{power}
\end{equation}%
These real power-law solutions for $H(t)$ exist iff $A^{2}\geqslant 8B$. For 
$\alpha ,\beta ,\gamma ,\Gamma \geqslant 0$ and $\gamma \neq \Gamma ,$ this
inequality is always satisfied. We can see this by defining 
\begin{equation}
\delta \equiv \frac{B}{A^{2}}=\frac{3(\alpha \Gamma +\beta \gamma +3\Gamma
\gamma )}{2(\alpha +\beta +3\gamma +3\Gamma )^{2}},  \label{delta}
\end{equation}%
so that $A^{2}\geqslant 8B$ iff $\delta \leqslant 1/8$. We see that the
denominator in equation (\ref{delta}) is always positive, so $\delta $ is
always non-singular and positive for finite and semi-definite positive
values of $\alpha ,\beta ,\gamma $ and $\Gamma $. It can also be seen that $%
\delta \rightarrow 0$ as either $\alpha $ or $\beta \rightarrow \infty $.
The maximum value of $\delta $ must therefore occur at finite values of $%
\alpha $ and $\beta $. If this maximum exists when $\alpha $ and $\beta $
are both non-zero then there must exist a point at which 
\begin{equation*}
\frac{\partial \delta }{\partial \alpha }=\frac{\partial \delta }{\partial
\beta }=0.
\end{equation*}%
Using (\ref{delta}) we can see that this condition is never met, so the
maximum value of $\delta $ must exist when $\alpha =0$, when $\beta =0,$ or
when $\alpha =\beta =0$. For $\alpha =0$ we will have a maximum at non-zero $%
\beta $ when 
\begin{equation*}
\left( \frac{\partial \delta }{\partial \beta }\right) _{\alpha =0}=0\qquad 
\text{and}\qquad \left( \frac{\partial ^{2}\delta }{\partial \beta ^{2}}%
\right) _{\alpha =0}\neq 0,
\end{equation*}%
which occurs iff $\beta =3(\gamma -\Gamma )$, for $\alpha $, $\beta $, $%
\gamma $, $\Gamma \geqslant 0$ and $\gamma \neq \Gamma $. Similarly, the
maximum can occur at non-zero $\alpha $ when $\beta =0$ iff $\alpha
=3(\Gamma -\gamma )$. We can choose, without loss of generality, $\Gamma
>\gamma $ so that the maximum value of $\delta $ occurs when $\alpha
=3(\Gamma -\gamma )$ and $\beta =0$, and we have the conclusion that 
\begin{equation*}
\delta \leqslant \delta _{max}=\frac{1}{8}
\end{equation*}%
for all $\gamma $, $\Gamma \geqslant 0$ and $\gamma \neq \Gamma $.

Having established the existence of the power-law solutions (\ref{power}),
we can now show that they behave as attractors of the general solution by
solving (\ref{H}). For $A^{2}>8B$ we find the solution 
\begin{equation}
H^{2}=a^{-A/2}(c_{1}a^{\sqrt{A^{2}-8B}/2}+c_{2}a^{-\sqrt{A^{2}-8B}/2}),
\label{Ha}
\end{equation}%
where $c_{1}$ and $c_{2}$ are constants. This solution to (\ref{master}) was
previously found by Chimento in \cite{Chim}. As $a\rightarrow \infty ,$ we
then have 
\begin{equation*}
H^{2}\rightarrow a^{-(A-\sqrt{A^{2}-8B})/2}
\end{equation*}%
and, as $a\rightarrow 0,$ 
\begin{equation*}
H^{2}\rightarrow a^{-(A+\sqrt{A^{2}-8B})/2}.
\end{equation*}%
These two equations can be integrated to obtain 
\begin{equation}
a_{\pm }\propto t^{(A\pm \sqrt{A^{2}-8B})/2B},  \label{power2}
\end{equation}%
which are the power-law solutions (\ref{power}), found earlier in eq. (\ref%
{h}). By integrating (\ref{Ha}) we can show explicitly the existence of the
above power-law attractors, and the smooth evolution of $a$ between them. It
is possible to integrate (\ref{Ha}) to get a solution in terms of $t$ and
the hypergeometric function $_{2}F_{1}(\tilde{a},\tilde{b};\tilde{c};x)$. An
expression in terms of more transparent functions can be found by defining a
new time coordinate $d\tau \equiv a^{-(A+\sqrt{A^{2}-8B})/4}dt$ and
integrating in terms of $\tau $. This gives the solution 
\begin{equation}
a\propto e^{\sqrt{c_{2}}(\tau -\tau _{0})}(1-e^{\sqrt{c_{2}}\sqrt{(A^{2}-8B)}%
(\tau -\tau _{0})})^{-2/\sqrt{(A^{2}-8B)}}.  \label{sol1}
\end{equation}
This is the same form for the evolution of $a$ that was found by Chimento
and Lazkoz in their investigation of phantom fluids in k-essence \cite{Chim2}%
. It can be seen from this expression that $a\rightarrow 0$ as $a\sim e^{%
\sqrt{c_{2}}\sqrt{(A^{2}-8B)}(\tau -\tau _{0})}$ when $\tau \rightarrow
-\infty $. In terms of the coordinate $t,$ this corresponds to the solution $%
a_{-}$ above. As $a\rightarrow \infty ,$ the solution smoothly approaches $%
a\sim (\tau -\tau _{0})^{-2/\sqrt{A^{2}-8B}}$ as $\tau \rightarrow \tau _{0}$%
, which corresponds to the solution $a_{+}$.

We have now shown that the two power-law solutions (\ref{power2}) exist and
for all $\alpha ,\beta ,\gamma, \Gamma \geqslant 0$ and are the attractors
of the smoothly evolving general solution at late and early times when $%
A^{2}>8B$.

It remains to investigate the limiting case $A^{2}=8B$. The exact solution
to equation (\ref{H}) when $A^{2}=8B$ is \cite{Chim} 
\begin{equation}
H^{2}=a^{-A/2}(c_{3}+c_{4}\ln a),  \label{sol2}
\end{equation}%
where $c_{3}$ and $c_{4}$ are constants. For $c_{4}=0$ this solution
corresponds to power-law expansion described by the degenerate case where $%
a_{+}\equiv a_{-}$. For $c_{4}\neq 0$ this solution is more complicated and
is bounded by $a=e^{-c_{3}/c_{4}}$ whilst approaching $H^{2}\sim a^{-A/2}\ln
a$ as $a\rightarrow 0$ or $\infty $, which does not describe a power-law
behaviour.

An illustrative special exact solution to equation (\ref{H}) exists when $%
B=A^{2}/9$, as was shown by Chimento \cite{Chim}. In this case (\ref{H}) can
be linearised to $\dddot{\psi}=0$ by the substitution $H=(3/A)\dot{\psi}%
/\psi $. Hence, for this special value of $B$ 
\begin{equation}
H=\frac{3(c_{5}+2c_{6}t)}{A(1+c_{5}t+c_{6}t^{2})},  \label{sol3}
\end{equation}%
where $c_{5},c_{6}$ are constants of integration. This expression can be
integrated to 
\begin{equation}
\left( \frac{a}{a_{0}}\right) ^{\frac{A}{3}}=1+c_{5}t+c_{6}t^{2},
\label{sol4}
\end{equation}%
where $a_{0}$ is a constant and the early and late time behaviour is clear
and has the same form as the power-law solutions (\ref{power}) when $%
B=A^{2}/9$.

\section{Evolution of the energy densities}

The conservation equations (\ref{2}) and (\ref{3}) can be used to construct
the second-order differential equation 
\begin{equation}
\frac{\rho ^{\prime \prime }}{\rho }+A\frac{\rho ^{\prime }}{\rho }+2B=0
\label{rho}
\end{equation}%
where $A$ and $B$ are defined as before and primes denote differentiation
with respect to the variable $\eta =\ln a$. This equation can be solved for $%
\rho $ and the corresponding solution for $\rho _{1}$ can then be found from
(\ref{3}). Substituting these solutions into the Friedmann equation (\ref{1}%
) gives, for $A^{2}>B$, the solution (\ref{Ha}) that was previously found by
solving the master equation (\ref{master}).

The advantage of considering the evolution of $\rho $ directly is that a
particularly interesting behaviour can be observed in the evolution of the
ratio $\rho /\rho _{1}$ for the self-similar solutions (\ref{h}). To find
this behaviour we first note that a solution to equation (\ref{rho}) is
given by 
\begin{equation*}
\rho =\rho _{0}a^{N}
\end{equation*}%
where $\rho _{0}$ is a constant and $2N=-A\pm \sqrt{A^{2}-8B}$. Substituting
this into equation (\ref{3}) gives the corresponding solution for $\rho _{1}$
\begin{equation*}
\rho _{1}=\rho _{10}a^{N}
\end{equation*}%
where $\rho _{10}=(N+3\gamma +\alpha )\rho _{0}/\beta $ is constant. These
solutions for $\rho $ and $\rho _{1}$, when substituted into the Friedmann
equation (\ref{1}), correspond to the self-similar solutions for $H$ given
by (\ref{h}). It is immediately apparent that $\rho $ and $\rho _{1}$ evolve
at the same rate and so the ratio $\rho /\rho _{1}$ is a constant quantity 
\begin{equation*}
\frac{\rho }{\rho _{1}}=\frac{\beta }{(N+3\gamma +\alpha )}
\end{equation*}%
during a period described by the power-law evolution (\ref{power2}). It is
this constant ratio in the energy density of two fluids with different
barotropic indices $\gamma $ and $\Gamma $ that has been used by a number of
authors in an attempt to alleviate the coincidence problem concerning the
the present-day values of the vacuum and matter energy-densities \cite%
{scal,free}.

\section{Three Examples}

The exact solutions found in the last sections provide us with extensions of
the analysis of several cosmological problems that have been studied in the
past, which can be defined by particular choices of the two parameters $A$
and $B$. As we have seen, the overall dynamical behaviour is determined by
the behaviour of the combination $\delta \equiv B/A^{2}.$

\subsection{Particle-antiparticle annihilation}

Consider the problem of the long-term evolution of a universe containing
equal numbers of electron-positron pairs \cite{ann}. If these particles are
assumed to be the lightest massive charged leptons then they cannot decay,
and can only disappear by means of the mutual annihilations $%
e^{-}e^{+}\rightarrow 2\gamma $. Page and McKee set up a model for the $%
e^{-}e^{+}$ annihilation into radiation that corresponds to taking the
special case $\alpha =0, \beta >0, \Gamma =1, \gamma =4/3$ in equations (\ref%
{2}) and (\ref{3}) and the definition of $\beta _{PM\text{ \ \ }}$by Page
and McKee is given in terms of our $\beta $ by $\beta \equiv 3\beta
_{PM}/(2-\beta _{PM}). $ They find the power-law solution with $%
h=h_{-}=2/(\beta +3)=(2-\beta _{PM)}/3$ which reduces to the usual dust FRW
model when $\beta =\beta _{PM}=0$ and there is no annihilation into
radiation. The effect of the annihilations is to push the expansion away
from the dust-dominated form with $a=t^{2/3}$ towards the radiation
dominated evolution with $a=t^{1/2}$. The other power-law solution
corresponds to the pure radiation case with $h=h_{+}=1/2.$ We can verify
that this power-law solution is an attractor by evaluating $\delta $, since
for the $e^{-}e^{+}\rightarrow 2\gamma $ annihilation $\beta _{PM}=(13-\sqrt{%
105})/8=0.3441$ so $\delta \equiv B/A^{2}=0.1247<1/8.$

\subsection{Primordial black-hole evolution}

A more complicated energy exchange problem was formulated by Barrow,
Copeland and Liddle \cite{pbh} who consider the problem of a power-law mass
spectrum of primordial black holes forming in the early universe and then
evolving under the effects of Hawking evaporation of the part of the mass
spectrum with Hawking lifetimes less than the expansion age. This has two
effects. The radiation background is supplemented by input from the
black-hole evaporation products and the fall in the total black-hole density
goes faster than the adiabatic $\rho _{bh}\propto a^{-3}$ that occurs in the
absence of decays because the black hole population is a pressureless gas to
a very good approximation, since $p/\rho \sim v^{2}\sim T/M_{bh}\sim
(m_{pl}/M_{bh})(t_{pl}/t)^{1/2}\approx 0$ for masses less than the Planck
mass $m_{pl}$ at times greater than the Planck time $t_{pl}$. Accretion of
background radiation in the radiation era of the universe by the black holes
could be included, but is negligible. This corresponds to our model in the
special case $\Gamma =1$, $\gamma =4/3,\alpha =0$ and 
\begin{equation*}
\beta =\frac{3(n-2)}{8-n},
\end{equation*}%
where the initial number density spectrum of black holes with masses between 
$m$ and $m+\delta m$ at time $t$ is given by $N(m,t)\propto m^{-n}$ and $%
n>2. $

A power-law solution was found in ref. \cite{ann} with $h=(8-n)/9$, so long
as $2<n<7/2$, and the black hole evaporations have a significant effect on
the expansion rate of the universe during the radiation era. We have 
\begin{eqnarray*}
A &=&\frac{2(25-2n)}{8-n}, \\
B &=&\frac{36}{8-n},
\end{eqnarray*}%
and so 
\begin{equation*}
\delta \equiv \frac{B}{A^{2}}=\frac{9(8-n)}{(25-2n)^{2}}.
\end{equation*}%
We see that the allowed range of $n$ $\in (2,7/2)$ corresponds to $\delta \
\in (6/49,1/8)$ and the expansion scale factor evolves as $a\propto
t^{(8-n)/9}.$ The $n=2$ limit corresponds to a pure dust-dominated
expansion, with $a\propto t^{2/3}$, while the $n=7/2$ limit corresponds to a
pure radiation-dominated evolution, with $a\propto t^{1/2}$. Again we see
that the power-law solution is an attractor for the general solution with $n$
in this range. When the expansion of the universe becomes dominated by cold
dark matter, with $\rho _{cdm}\propto a^{-3}$, a power-law scaling solution
no longer exists because the radiation products from the black-hole
evaporations now make a negligible contribution to the total density of the
universe, which is dominated by $\rho _{cdm}>\rho _{bh}>>\rho _{\gamma }$,
and $a\propto t^{2/3}$ becomes the attractor for the evolution of the
expansion scale factor.

\subsection{Vacuum decay}

The cosmological evolution created by the decay of a vacuum stress ($\rho
_{1}=\rho _{v})$ into equilibrium radiation was considered by Freese et. al.
and many other authors \cite{free}. It is described by a special case of our
equations (\ref{2}) and (\ref{3}) with $\Gamma =0,\gamma =4/3,\alpha =0$ and 
$\beta >0$. It represents the decay of a scalar field stress with $p\approx
-\rho $ into radiation. In this case we have

\begin{eqnarray}
A &=&\beta +4,  \notag \\
B &=&2\beta ,  \notag \\
\delta &\equiv &\frac{B}{A^{2}}=\frac{2\beta }{(\beta +4)^{2}},  \label{v2}
\end{eqnarray}%
with $h_{+}=1/2$ and $h_{-}=2/\beta $. We see that the first of these
corresponds to the degenerate situation with pure radiation. The second
solution has $a\propto t^{2/\beta }$ and requires $\beta >3$ if the
evolution of the universe is to have a matter-dominated era following a
radiation era. As the value $\beta $ increases, the dominance of the vacuum
contribution slows the expansion whereas in the limit $\beta \rightarrow 0$
the expansion rate increases without bound and the dynamics approach the
usual vacuum-energy dominated de Sitter expansion with $a\propto \exp (t%
\sqrt{\rho _{v}}/3).$ Again we see that this simple solution can be
generalised by using the full analysis provided above. We see from (\ref{v2}%
) that we always have $\delta \leqslant 1/8$ with the maximum of $\delta $
achieved when $\beta =4$. The solution (\ref{sol3}) - (\ref{sol4}) arises
for $\delta =1/9$ which occurs when $\beta =2$ or $\beta =8.$

\section{Anti-Decaying Fluids}

\label{antidecay}

It was shown in section \ref{decaysec} that for $\alpha $, $\beta $, $\gamma 
$, $\Gamma \geqslant 0$ and $\gamma \neq \Gamma $ the maximum value that $%
\delta $ can take is $1/8$. If we relax these assumptions, then $\delta $
can take values greater than $1/8$ and the qualitative character of the
solutions to (\ref{H}) is significantly altered. We will begin by
investigating the conditions required for $\delta >1/8$.

In the previous section it was shown that for $\alpha $, $\beta $, $\gamma $%
, $\Gamma \geqslant 0$ and $\gamma \neq \Gamma $ the only point at which $%
\delta =1/8$ is at $\alpha =3(\Gamma -\gamma )$ and $\beta =0,$ and at all
other points in this parameter range we have $\delta <1/8$. It can be seen
from (\ref{delta}) that $\delta =1/8$ when 
\begin{equation*}
\alpha =\left( \sqrt{3(\Gamma -\gamma )}\pm \sqrt{-\beta }\right)
^{2}\geqslant 0
\end{equation*}%
and that the first derivatives of $\delta $ are non-zero at any point where
this condition is satisfied. These values of $\alpha $ therefore separate
regions where $\delta <1/8$ from those where $\delta >1/8$. It can also be
seen that $\delta >1/8$ only if $\alpha >0$ and $\beta <0$. These conditions
correspond to the fluid $\rho $ decaying and the fluid $\rho _{1}$
anti-decaying. (By `anti-decaying' we mean gaining energy in proportion to
its energy density, instead of losing it). An example of an anti-decaying
fluid is a ghost field which radiates away energy; here the energy density
of the ghost is negative, so a negative value of $\beta$ is required for the
radiation to carry away energy.

\begin{figure}[t]
\centering \includegraphics{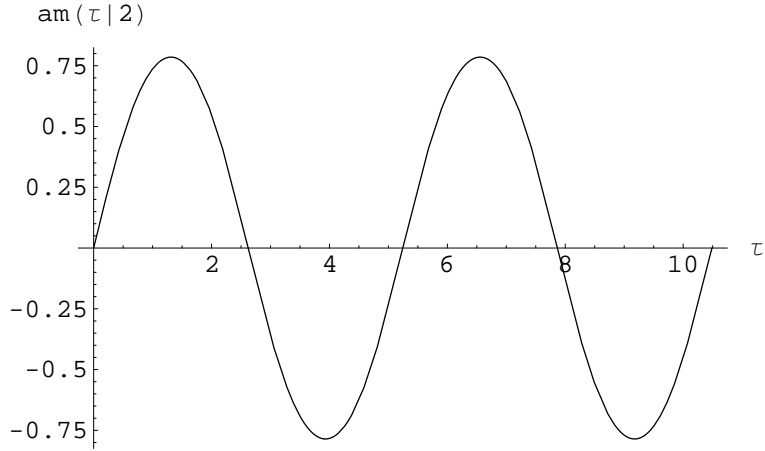}
\caption{The Jacobi amplitude am$(\protect\tau |2)$}
\label{jacobi}
\end{figure}

For $\delta >1/8$ the exact solution to equation (\ref{Ha}) is 
\begin{equation}
H^{2}=a^{-A/2}\cos \left( \frac{1}{2}\sqrt{8B-A^{2}}\ln a\right) ,
\label{Ha2}
\end{equation}%
where integration constants have been rescaled into $a$ and $H$. Again, this
equation is difficult to solve in term of the coordinate $t$. By introducing
the new coordinate $d\tau \equiv a^{A/4}dt,$ we get 
\begin{equation*}
\frac{d\ln a}{d\tau }=\cos \left( \frac{1}{2}\sqrt{8B-A^{2}}\ln a\right) ,
\end{equation*}%
which can be integrated to obtain a closed form for the expansion scale
factor: 
\begin{equation}
a(\tau )=\exp \left\{ \frac{4}{\sqrt{8B-A^{2}}}\;\;\text{am}\left( \frac{%
\sqrt{8B-A^{2}}}{4}(\tau -\tau _{0})\;\vline\;2\right) \right\} ,
\label{sol5}
\end{equation}%
where $\tau _{0}$ is constant and am$(\hat{a}|\hat{b})$ is the Jacobi
amplitude, shown in figure \ref{jacobi}. The form of $a(\tau )$ in (\ref%
{sol5}) is an always-positive oscillatory function of the time $\tau $, with
constant amplitude. The corresponding solution in terms of the coordinate $t$
will therefore also be oscillatory with constant amplitude. 

Whilst in the previous section we found that, for $\delta <1/8,$ the scale
factor evolves as a smooth function with early and late-time power-law
behaviour, we have found for $\delta >1/8$ a substantially different
behaviour. The scale factor now oscillates in time and does not display the
simple power-law behaviour found in the $\delta \leqslant 1/8$ situations.

\section{Phantom Fluids}

We have so far only discussed the cases where $\gamma $, $\Gamma \geqslant 0$
and $\gamma \neq \Gamma $. This assumption is useful as it means that $%
\delta $ is non-singular in the parameter range $\alpha $, $\beta \geqslant
0 $, for which it was shown in section \ref{decaysec} that the maximum value
of $\delta $ is $1/8$. This result was subsequently used in section \ref%
{antidecay} to show that there exists a parameter range with $\alpha >0$ and 
$\beta <0$ for which $\delta >1/8$. In this section we will relax the
positive semi-definite assumption on the parameters $\gamma $ and $\Gamma $,
extending the analysis we have so far performed to the case of so called
`phantom' fluids. We begin by showing that in the parameter range $\alpha $, 
$\beta >0$ there exist no points at which $\delta \rightarrow +\infty $. For
this to occur we would require the simultaneous satisfaction of the
conditions 
\begin{equation*}
\alpha \Gamma +\beta \gamma +3\Gamma \gamma >0\qquad \text{and}\qquad \alpha
+\beta +3\gamma +3\Gamma =0.
\end{equation*}%
Using the second of these conditions, we can eliminate $\alpha $ in the
first to find 
\begin{equation*}
\beta (\gamma -\Gamma )-3\Gamma ^{2}>0\qquad \text{or}\qquad \beta <-\frac{%
3\Gamma ^{2}}{(\Gamma -\gamma )}\leqslant 0,
\end{equation*}%
where we still assume $\Gamma >\gamma $, without loss of generality.
Similarly, we can obtain for $\alpha $ the expression 
\begin{equation*}
\alpha >\frac{3\gamma ^{2}}{(\Gamma -\gamma )}\geqslant 0.
\end{equation*}%
These two inequalities show that $\delta \rightarrow +\infty $ can only
occur in the parameter space $\alpha >0$ and $\beta <0$. Therefore, in the
range $\alpha $, $\beta >0$ the only singularities in $\delta $ that can
occur are those in which $\delta \rightarrow -\infty $. In this case we can
again show, using the arguments in section \ref{decaysec}, that the maximum
value of $\delta $ when $\alpha $, $\beta \geqslant 0$ is $1/8$. The
argument showing the existence of a region where $\delta >1/8$ in the range $%
\alpha >0$ and $\beta <0$ now follows in exactly the same way as for the $%
\gamma $, $\Gamma \geqslant 0$ case, given in section \ref{antidecay}.

The form of the solutions in the regions where $\delta <1/8$ and $\delta
>1/8 $ are the same as in the non-phantom case, and are given by equations (%
\ref{Ha}) and (\ref{Ha2}).

\section{Discussion}

We have determined the general solution of a simple model with the exchange
of energy between two fluids in an expanding Friedmann universe of zero
spatial curvature. The total energy of the exchange is conserved and the
model allows energy inputs and outflows proportional to the densities of the
two fluids. A number of simple examples of this sort already exist, such as
particle decays or particle-antiparticle annihilations into radiation,
particle production, the evaporation of a population of primordial black
holes, the decay of a cosmological vacuum or cosmological `constant', and
energy exchanges between quintessence and ordinary matter or radiation.
However, these examples are restricted to one-way energy exchange and do not
prove that the scaling solutions that they employ are attractors for the
general solution. We have established the existence and form of simple
power-law solutions for the expansion scale factor in the case of two-way
energy exchange between fluids and determined that they are attractors for
the late-time evolution in situations that are usually regarded as generic.
If we allow one fluid to be anti-decaying then we can move into a domain
where these power-law solutions are no longer attractors. Again, we find the
general behaviour for these cosmologies. These solutions provide a simple
model for the study of a wide range of energy exchange problems in cosmology
and also reveal the conditions under which power-law solutions previously
used to solve some of these problems are stable attractors. They provide a
simple model for many future studies of a variety of interacting fluid
cosmologies.

\section{Acknowledgements} We would like to thank Luis Chimento for helpful
references. T. Clifton acknowledges support from PPARC.

\end{document}